# A Coarse-Grained Biophysical Model of *E. coli* and Its Application to Perturbation of the rRNA Operon Copy Number

Arbel D. Tadmor , Tsvi Tlusty

Department of Physics of Complex Systems, Weizmann Institute of Science, Rehovot, Israel

## Abstract

We propose a biophysical model of *Escherichia coli* that predicts growth rate and an effective cellular composition from an effective, coarse-grained representation of its genome. We assume that *E. coli* is in a state of balanced exponential steady-state growth, growing in a temporally and spatially constant environment, rich in resources. We apply this model to a series of past measurements, where the growth rate and rRNA-to-protein ratio have been measured for seven *E. coli* strains with an rRNA operon copy number ranging from one to seven (the wild-type copy number). These experiments show that growth rate markedly decreases for strains with fewer than six copies. Using the model, we were able to reproduce these measurements. We show that the model that best fits these data suggests that the volume fraction of macromolecules inside *E. coli* is not fixed when the rRNA operon copy number is varied. Moreover, the model predicts that increasing the copy number beyond seven results in a cytoplasm densely packed with ribosomes and proteins. Assuming that under such overcrowded conditions prolonged diffusion times tend to weaken binding affinities, the model predicts that growth rate will not increase substantially beyond the wild-type growth rate, as indicated by other experiments. Our model therefore suggests that changing the rRNA operon copy number of wild-type *E. coli* cells growing in a constant rich environment does not substantially increase their growth rate. Other observations regarding strains with an altered rRNA operon copy number, such as nucleoid compaction and the rRNA operon feedback response, appear to be qualitatively consistent with this model. In addition, we discuss possible design principles suggested by the model and propose further experiments to test its validity.

## Introduction

The rRNA (*rrn*) operons of *E. coli* have the important role of determining ribosome synthesis in the cell (c.f. [1–6] for reviews). These operons are unique in the sense that a wild-type (WT) *E. coli* cell carries seven copies of this operon per chromosome [7] (other bacteria have copy numbers ranging between 1 and 15 [8]). This copy number also appears to be selectively maintained. *S. typhimurium* for example, from which *E. coli* is thought to have diverged 120–160 million years ago [9], also has seven copies of this operon [10] and in evolution experiments of up to $10^4$ generations no deviations from the WT copy number have been observed [11]. These findings raise the question of what underlying mechanisms, if any at all, fixed this copy number to be seven and not six or eight. In other cases it has been shown that the WT genome configuration maximizes fitness [12–14]. Thus, can it be shown that this copy number maximizes fitness?

In general, this question is hard to answer because the natural environment of *E. coli* is expected to vary both spatially and temporally [15,16], thereby invoking complex physiological responses in the cell that are complicated to model. We therefore consider a much simpler scenario, where a resource-rich environment is spatially and temporally constant, and where the cell is at a state of balanced exponential steady-state growth [3], such that it has a well defined and reproducible growth rate and physiological state [17,18]. In such an environment, cells that can outcompete their rivals will takeover the population and thus fix their genotype. Thus we refer to fitness in the narrow sense that previous authors have used [11,14], i.e. it is the potential capacity of a cell in exponential growth to outcompete another strain population wise. Therefore, for an exponentially growing cell in a constant and rich environment, fitness would be, by definition, the growth rate of that cell.

Experimental evidence indeed suggests that for exponentially growing cells, cells with altered rRNA operon copy numbers have a lower growth rate. In a series of experiments, the Squires group has measured the growth rate and cell composition of seven strains of *E. coli* with rRNA operon copy numbers ranging from one to seven copies per chromosome [19]. All strains were grown in the same nutrient rich environment and measurements were performed on cells in exponential phase. These experiments show that cells with fewer than six rRNA operons have a considerable lower growth rate [19] (c.f. Figure 2A presented later in the text). For example, cells with five functional rRNA operons have a 21%

## Author Summary

A bacterium like *E. coli* can be thought of as a self-replicating factory, where inventory synthesis, degradation, and management is concerted according to a well-defined set of rules encoded in the organism's genome. Since the organism's survival depends on this set of rules, these rules were most likely optimized by evolution. Therefore, by writing down these rules, what could one learn about *Escherichia coli*? We examined *E. coli* growing in the simplest imaginable environment, one constant in space and time and rich in resources, and attempted to identify the rules that relate the genome to the cell composition and self-replication time. With more than 4,400 genes, a full-scale model would be prohibitively complicated, and therefore we "coarse-grained" *E. coli* by lumping together genes and proteins of similar function. We used this model to examine measurements of strains with reduced copy number of ribosomal-RNA genes, and also to show that increasing this copy number overcrowds the cell with ribosomes and proteins. As a result, there appears to be an optimum copy number with respect to the wild-type genome, in agreement with observation. We hope that this model will improve and further challenge our understanding of bacterial physiology, also in more complicated environments.

lower growth rate than WT cells, while cells with only one functional rRNA operon have a 50% lower growth rate than WT cells [19]. In addition, a strain carrying extra rRNA operons on a plasmid exhibited a 22% reduction in growth rate relative to a WT control strain with a plasmid expressing nonfunctional rRNA [20].

To gain further insight into these findings, we sought to formulate a model of *E. coli* that could predict phenotype, such as growth rate and cell composition, directly from DNA related parameters, such as the rRNA operon copy number, while keeping the complexity of the model to a minimum. The model of *E. coli* proposed here differs from existing models of *E. coli* in several respects. Traditionally, *E. coli* has been monitored in different or changing environments [17,21–23], and existing models have attempted to predict *E. coli*'s response to such environmental perturbations [12,23–26]. However, since disparate environments are expected to induce disparate genetic networks, we anticipate that such a strong perturbation will be difficult to capture in a simple model that attempts to predict phenotype from DNA related parameters (c.f. S2.3 in Text S1). Existing models of *E. coli* tend to fall into two classes. One class includes very complex models, involving tens to hundreds of equations [12,24,25], which do not lend themselves to simple interpretation. The other class involves simple and elegant models of *E. coli* that followed the Copenhagen school [22,23,26,27] (see [17] for review). These classic models, however, do not relate genome to growth rate and composition, nor do they make reference to certain key physical processes in the cell elucidated since. Included in our current model are the relationships of genome to growth rate and cell composition, reflecting key physical processes now better understood, such as RNA polymerase (RNAp)-promoter interaction [3,28], RNAp autoregulation [29], ribosome-ribosome binding site (RBS) interaction [30–33], mRNA degradation [34–40], DNA replication initiation [41–43] and macromolecular crowding (see below) [44–51]. In addition, we have attempted to find the middle ground in terms of complexity by coarse-graining, for simplicity, certain features of the cell: in the spirit of previous works [28,52], the genome has been lumped into a small set of "gene classes" that represent all transcription and translation within the cell for the given environment. Similarly, the cell composition was reduced to a small set of variables accounting for the macromolecule content of the cell. The resulting type of model is referred to as a *Coarse-Grained Genetic Reactor* (CGGR).

Another point of difference with respect to existing models of *E. coli* is that in this model we take into account possible global biophysical effects resulting from the high volume fraction of macromolecules in *E. coli*, a state commonly termed "crowding" [45]. Formulating such a biophysical model for *E. coli* raises the basic question: is the macromolecular volume fraction, $\Phi = V_{macro}/V_{cell}$, inside *E. coli* constrained to be fixed or does it change for genetically perturbed cells? We have explored both of these possibilities in what we refer to as the *constrained* ($\Phi = $ const) and *unconstrained* ($\Phi \neq $ const) CGGR models.

Using the CGGR modeling approach, we have modeled the seven strains engineered by the Squires group and have calculated their growth rate and their effective cellular composition. We were able to reproduce the experimental data within a model in which macromolecular volume fraction was allowed to change for genetically perturbed cells. These findings, along with other biological considerations, seem to favor the unconstrained CGGR model (see Discussion). According to this model, increasing the chromosomal rRNA operon copy number beyond seven will overcrowd the cytoplasm with ribosomes and proteins. Under such over-crowded conditions, we expect that binding affinities will weaken due to prolonged diffusion times. As a result, given this assumption, we show that the growth rate of an exponentially growing cell in a constant rich medium will not increase substantially beyond its WT growth rate when the rRNA operon copy number is increased beyond seven. Although we have not shown that the maximum in growth rate is a global maximum, since we only perturbed one genetic parameter, this result suggests that—at least for the case of a cell undergoing balanced exponential steady-state growth in a constant and rich medium—basic kinetic and biophysical considerations may have an important role in determining an optimal rRNA operon copy number (see Discussion).

Besides explaining the Squires data, the unconstrained CGGR model is qualitatively consistent with observations regarding nucleoid compaction in the inactivation strains and with the *rrn* feedback response originally observed by Nomura and coworkers (see Methods and Discussion). Thus, the CGGR model may offer an initial conceptual framework for thinking about *E. coli* as a whole system at least for the simplified environment considered. More complex genetic networks may subsequently be embedded into this model enabling one to analyze them in the larger, whole cell framework. Such a model may also help elucidate how *E. coli* works on a global scale by making experimentally testable predictions and suggesting experiments (see Discussion). We will also consider possible insights into the "design principles" of *E. coli* suggested by the CGGR model, such as intrinsic efficiency of resource allocation and decoupling of DNA replication regulatory mechanisms from cell composition.

## Methods

### The Cell as a Coarse-Grained Genetic Reactor (CGGR)

Our goal is to formulate a model of *E. coli* that predicts phenotype, such as growth rate and the cell composition, from parameters directly related to the genome, while keeping complexity to a minimum. To reduce the complexity of this problem we coarse-grained both input parameters (the genome) and output parameters (the cell composition and growth rate). The genome (input) was lumped into four basic *gene classes*: RNA polymerase (RNAp), ribosomal protein (r-protein), stable RNA and bulk protein. The gene classes are represented by *genetic parameters* such as: genetic map locations, promoter strengths, RBS strengths, mRNA half-lives and transcription and translation times, all of which can be, in principle, linked to the DNA. Genetic parameters have been determined based on empirical data for the WT growth rate (or very close to it) and represent all transcription and translation within the cell at that growth rate (see Results for more details).

The cell composition (output) was reduced to the following five macromolecule classes: free functional RNAp, total RNAp, free functional ribosomes, total ribosomes and bulk protein. The bulk protein class represents all other cell building and maintenance proteins in the cell [53]. The concentration of these macromolecules, together with the growth rate constitutes six state variables that define the *cell state*. Table 1 gives an example of the observed WT cell state at 2.5 doub/h. In the Discussion we consider the applicability of this choice of coarse-graining.

## The Feedback Mechanisms within a Coarse-Grained Model of *E. coli*

After coarse-graining the cell, one can map the various feedback mechanisms that exist between these coarse-grained components, as illustrated in Figure 1A. Transcription of the various gene classes by RNAp [29] is depicted on the left, and translation of mRNA by ribosomes on the right. Ribosomes are shown to be assembled by combining rRNA with r-protein. r-proteins synthesis rate is regulated to match the rRNA synthesis rate [1] as indicated by the black arrow in Figure 1A.

RNAp naturally has positive feedback to all promoters, and ribosomes have positive feedback to all RBS. In the case of RNAp, it has been shown [29] that the $\beta\beta'$ subunits, which limit the production of RNAp (c.f. discussion in [17]), repress their own translation, and the functional, assembled RNAp holoenzyme represses transcription of $\beta\beta'$. While the details of the RNAp autoregulation are still being elucidated, the latter finding suggests that the apparent fast response of the negative translational autoregulation of the $\beta\beta'$ operon keeps the concentration of total RNAp fixed, at least approximately (for a detailed discussion see S2.1 in Text S1). The level of RNAp may also be modulated by guanosine 5′-diphosphate 3′-diphosphate (ppGpp) [3,17,28], however, since ppGpp levels were measured to be constant for strains with increased or decreased number of rRNA operons [20,54], this modulation will not be relevant in this analysis (see discussion below).

Finally, there is the feedback arising from the translation-degradation coupling, indicated in Figure 1A by the dashed green line connecting ribosomes to mRNA degradation. Ribosomes bound to the RBS of mRNA protect the mRNA from degradation by preventing RNase E – thought to be the primary endonuclease

**Table 1.** The cell state at $\mu = 2.5$ doub/h, 37°C.

| State variable | $\mu$ (doub/h) | $n_{bulk}$ | $n_{RNAp}$ | $n_{RNAp,free}$ | $n_{ribo}$ | $n_{ribo,free}$ |
| --- | --- | --- | --- | --- | --- | --- |
| Measured value | 2.5 | $5.76 \cdot 10^6$ | 11400 | 890 | 72000 | 4700 |

Cell composition ($n_i$) was either directly measured or estimated from empirical data and is given in units of molec/cell. Measurement error is expected to be around 15%, mostly due to culture-to-culture variation [17]. More details can be found in Table S2.

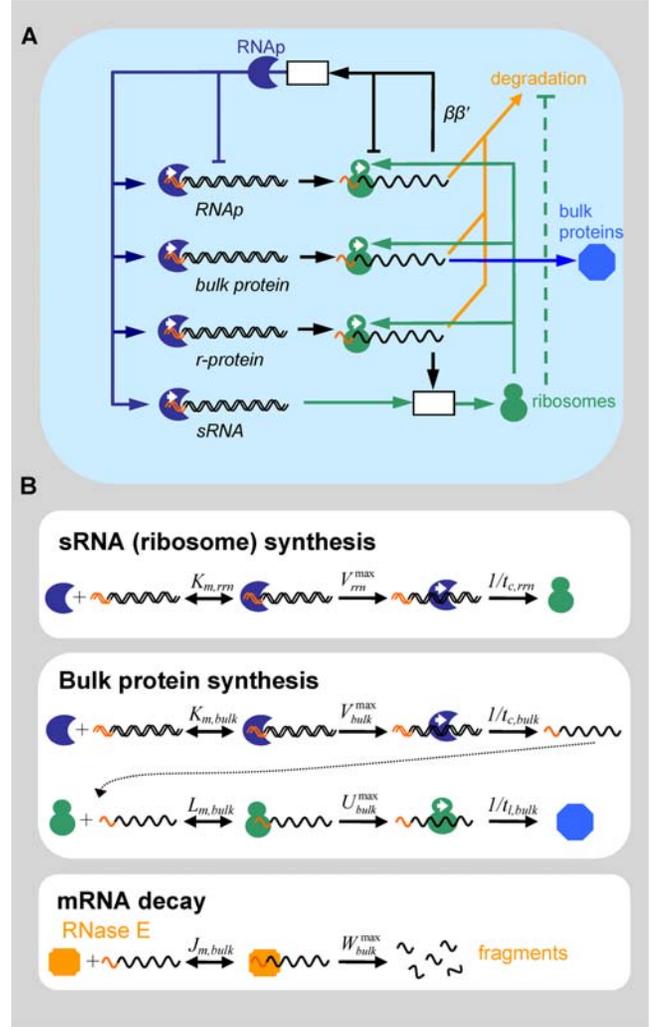

**Figure 1. The coarse-grained genetic reactor (CGGR) model of *E. coli*.** (A) A schematic diagram of the cell as a CGGR. This figure depicts the various gene classes (double stranded DNA on the left), mRNA (single stranded RNA on the right) and their expressed products, stable RNA and the various positive (→) and negative (⊣) feedbacks between these components. The colored tips on the DNA and mRNA represent the promoter binding sites and the ribosome binding sites (RBSs) respectively. The white boxes denote the process of assembly of functional complexes from immature subunits. The light blue background represents the finite volume of the cell in which reactions take place, which is determined by the DNA replication initiation system (see text and S2.2 in Text S1). (B) The basic reactions taking place in the cell: $K_{m,i}$, $V_i^{max}$ are the Michaelis-Menten (MM) parameters for transcription initiation of gene class i [3,28], followed by transcription at a rate of $1/t_{c,i}$. $L_{m,i}$, $U_i^{max}$ are the MM parameters of gene class i for translation initiation (i.e. binding of a 30S ribosome subunit to a RBS) [30,31], followed by translation at a rate of $1/t_{l,i}$. Note that in this scheme, the 30S-RBS binding affinity, $L_{m,i}$ includes the 30S interaction with the secondary structure of mRNA [30]. mRNA degradation is primarily achieved via the endonuclease RNase E [35] and is assumed to follow MM like kinetics ($J_{m,i}$, $W_i^{max}$) [36,37]. In the scheme considered here, RNase E competes with the 30S subunits in binding to a vacant RBS [38,39], functionally inactivating it when it binds [38]. This results in a coupling between translation and degradation [34]. Delays due to assembly are incorporated separately. The MM parameters, together with the time constants and map locations define *genetic parameters* for the various gene classes, which can be easily "tweaked" by base pair mutations (c.f. Table 2 and Table S1).

in *E. coli* [35,55] – from binding to the 5′ end of the mRNA and cleaving it (for recent reviews see [34,38,39]). RNase E is considered to be the partially or fully rate-determining step in the mRNA degradation process [35,55–59]. Here we have modeled this effect by allowing only mRNA's with a vacant RBS to be cleaved [38,39,60]. This feedback manifests itself as dependence of mRNA half-life on the probability that the RBS is vacant, as suggested by observation [34,56,61] (see S2.7 in Text S1 and discussion further below).

We shall refer to the feedbacks depicted in Figure 1A as *internal feedbacks*. If the macromolecular volume fraction is allowed to change, then an additional internal feedback arises due to the fact that the binding affinities of RNAp and ribosomes to their corresponding binding sites may change due to crowding effects. This kind of dependence on the crowding state of the cell offers an additional feedback path not explicit in Figure 1A. We will elaborate on this point in the Results section. Also not explicit in Figure 1A is DNA replication that determines gene concentration. This issue will be further discussed below.

## The Feedback Response of the rRNA Operons

It has long been known that there is some form of feedback control on rRNA operons that responds to any artificial attempt to manipulate ribosome synthesis [1–6,20,62], yet the source of this feedback has remained controversial. Since we will be considering perturbations on the rRNA operon copy number, which affect ribosome synthesis, it is pertinent to identify any additional effectors that apply a feedback within the system.

Nomura and his coworkers noted that cells with increased number of rRNA operons did not exhibit a significant increase in rRNA transcription [62], i.e. the transcription per rRNA operon decreased by means of some feedback. Furthermore, the absence of this feedback in cells overproducing nonfunctional rRNA, and the observation of a feedback response in strains in which ribosome assembly was blocked, suggests that complete ribosomes are involved in the feedback response [62]. This became known as the "ribosome feedback regulation model" (c.f. discussions in [1,3,5]). Direct effect of ribosomes on rRNA transcription could not, however, be observed *in vitro* [62], and it was suggested by these authors that this regulation may be achieved indirectly [62]. Further experiments indicated that the feedback depends on translating ribosomes (or translational capacity) rather than free ribosomes [1,63]. Later studies have demonstrated the feedback response for various other perturbations that attempted to artificially manipulate ribosome synthesis rate, including: increasing rRNA operon copy number [20,64,65], decreasing rRNA operon copy number [54], overexpressing rRNA from an inducible promoter [66], deleting the *fis* gene [20,67] (see below), muting the *rpoA* gene coding for the α subunit of RNAp [20,68] and more (c.f. [2,20]). Since many of these perturbations [20], as well as perturbations in nutritional conditions [2,69], correlated with changes in the concentration of ppGpp and nucleoside triphosphate (NTP), Gourse and his coworkers have proposed that NTP and ppGpp are the feedback regulators [6,69]. In addition, these authors have suggested a model where translating ribosomes consume or generate NTP and ppGpp and thus are able to achieve homeostasis of rRNA expression on a rapid time scale [6,69]. Yet these authors also point out that these effectors cannot explain the feedback response specifically associated with changes in rRNA gene dosage [64] (the perturbations considered in this study). In this case, it has been demonstrated that the small molecule ppGpp has no effect on rRNA synthesis rate both in the case where rRNA gene dosage was increased [20] or decreased [54] since ppGpp concentration remains constant in these strains (also indicating that tRNA imbalance was not a problem in those strains). In addition, feedback inhibition due to increased rRNA gene dosage was of the same magnitude in both wild-type cells and strains lacking ppGpp [70]. Similarly, the concentration of the small effector NTP was shown to be constant when decreasing or increasing the rRNA gene dosage [64]. In a different study, NTP concentration decreased by only a small amount (20%) when rRNA gene dosage was increased [20], such that those authors concluded that the small change in NTP concentration appears to be insufficient to account for the entire effect on transcription initiation. Due to these findings, Gourse and coworkers concluded that there may be additional mediators involved in feedback control of rRNA expression when altering the rRNA operon gene dosage [2,20]. We show that internal feedbacks may account, at least partially, for the feedback response, although an additional effector may still be involved. In the Discussion we analyze model predictions and compare them to observations regarding this effect. We will also discuss the predicted feedback response in the context of Nomura and coworkers' feedback model and show that there appears to be no contradiction between the two.

## Additional Factors Affecting rRNA Expression

In addition to the small molecules mentioned above, rRNA transcription is further modulated by transcription factors like Fis, HN-S and DskA, as well as the UP element [1,2,4–6,69,71], however there is currently no experimental evidence to suggest that these factors are linked to the feedback response to altered rRNA gene dosage. DskA, for example, a small molecule that binds to RNAp, is thought to amplify effects of small nucleotide effectors such as ppGpp and NTP [4,6,72]. DskA concentration, however, was found to be unchanged with growth rate and growth phase and therefore it apparently does not confer a novel type of regulation on rRNA synthesis [3,72] and is thus considered to be a co-regulator rather than a direct regulator [4]. Fis stimulates rRNA transcription by helping recruit RNAp to the promoter through direct contact with the α subunit of RNAp, while the UP element, a sequence upstream of the promoter, binds the α subunit of RNAp and can greatly stimulate rRNA transcription [2,4–6,71]. Although Fis levels change throughout the growth cycle [4,5], strains lacking Fis binding sites retain their regulatory properties [2,5,67] indicating that *fis* is not essential for regulation of rRNA transcription during steady-state growth [67], and perhaps just plays a role in control during nutritional shift-ups and onset of the stationary phase [5]. HN-S concentration changes with the growth phase of an *E. coli* culture [2,69,71] and is thought to be associated with regulation related to stress [15], particularly in stationary phase [4,69]. Since there is currently no direct evidence that shows that any of these or other factors are associated with the feedback response to rRNA gene dosage perturbation, no such factors were included in the proposed models, yet future experiments may prove otherwise (c.f. Discussion).

## Kinetic Equations

The biochemical reactions that make up the feedback network illustrated in Figure 1A are approximated, for simplicity, by Michaelis-Menten type kinetics [3,28,30,36,37], as is illustrated in Figure 1B. These reactions include: stable RNA synthesis, bulk protein synthesis and bulk mRNA decay. Since ribosomes and the bulk of proteins in *E. coli* are stable on timescales of several generations [35,73], their degradation can be neglected compared to the fast doubling time of the cell (~30 min). We also do not need to explicitly consider r-protein synthesis since ribosome synthesis is limited by rRNA [1]. Finally we note that the free RNAp in these reactions may include RNAp bound nonspecifically to DNA and in rapid equilibrium with it [28] that may locate promoters by a type of 1-D sliding mechanism [74]. In the current model, all inactive RNAp was assumed to be associated with pause genes (c.f. [28] and e.g. Table S4) and thus inaccessible to promoters. However, it may be that some of these inactive RNAp molecules are just bound nonspecifically to the DNA [28] perhaps serving as an additional reservoir of RNAp.

Since the Squires strains were measured under steady-state conditions, we consider next the steady-state equations implied by Figure 1B.

## The CGGR Steady-State Equations

The reactions in Figure 1B can be readily expressed as rate equations and analyzed at steady-state. Although the full derivation is rather lengthy (see S2.5 in Text S1), the final equations lend themselves to simple interpretation. The average transcription [3,75] and translation [30,31,76] initiation rates are given by the usual Michaelis-Menten relations

$$V_i = V_i^{\max} \frac{1}{1+K_{m,i}/n_{RNAp,free}}, U_i = U_i^{\max} \frac{1}{1+L_{m,i}/n_{ribo,free}} \quad (1)$$

where $n_i$ denotes the concentration of species $i$, $V_i^{\max}$ and $U_i^{\max}$ are the maximum transcription and translation initiation rates of the $i$-th gene class respectively, and $K_{m,i}$ and $L_{m,i}$ are RNAp holoenzyme and 30S ribosome subunit binding affinities of the $i$-th gene class to their corresponding binding sites respectively, measured in units of concentration (see Table 2 for notation list and units). Using this notation, the RNA transcript synthesis rate per unit volume is $v_i = d_i V_i$ (where $d_i$ is the gene concentration of the $i$-th gene class) and the number of translations per mRNA is $u_i = U_i T_{1/2,i}^{fun}/\ln 2$, where $T_{1/2,i}^{fun}$ is the functional half-life of the $i$-th gene class mRNA. Therefore, the protein synthesis rate per unit volume of gene class $i$ is $v_i u_i$. In this notation, the five equations of state take the form:

$$\begin{aligned}
&\text{(i) } n_{RNAp} \approx \text{const} \\
&\text{(ii) } n_{bulk} = \frac{1}{\alpha} v_{bulk} u_{bulk} \\
&\text{(iii) } n_{ribo} = \frac{1}{\alpha} v_{rrn} \\
&\text{(iv) } n_{RNAp} = n_{RNAp,free} \\
&\quad + (t_{c,bulk} v_{bulk} + t_{c,r-protein} v_{r-protein} + t_{c,rrn} v_{rrn}) \\
&\quad + (1-e^{-\alpha \tau_{RNAp}}) n_{RNAp} \\
&\text{(v) } n_{ribo} = n_{ribo,free} + (t_{l,bulk} v_{bulk} u_{bulk} + t_{l,r-protein} v_{rrn}) \\
&\quad + (1-e^{-\alpha \tau_{ribo}}) n_{ribo}
\end{aligned} \quad (2)$$

where $t_{c,i}$ and $t_{l,i}$ are the times to transcribe and translate the $i$-th gene class respectively, and $\tau_i$ is the average assembly time for component $i$ (the boxes in Figure 1A). Equation (i) states that the total RNAp concentration is constant. This is due to our assumption that the negative autoregulation of RNAp is ideal. This somewhat naïve model for the autoregulation of RNAp can be, in principle, replaced with a more sophisticated model describing the steady-state response of the negative transcriptional and translational autoregulation of RNAp, once the details of this mechanism are known. Equations (ii) and (iii) are the bulk protein and ribosome synthesis equations respectively, assuming exponential growth, i.e. dilution at a rate of $\alpha = \mu \ln 2$, where $\mu$ is the doubling rate. Note that $v_{rrn}$ is the total ribosome synthesis rate per unit volume. Finally, (iv) and (v) are conservation equations for RNAp and ribosomes within the cell. In Eq. (iv), these terms include (left to right): free RNAp, bound RNAp and immature RNAp (a modified version of Eq. (iv) was first derived in [28]). Similar terms exist in the ribosome conservation equation (v). The contribution of RNAp to the conservation equations was neglected since it constitutes less than 2% of the total protein mass [17]. Note that in the second term of (v), the number of bound ribosomes to the r-protein class is determined by the time it takes to translate all r-proteins and the total *rrn transcription rate*, due to the matching of r-protein synthesis rate and rRNA synthesis rate through regulation at the r-protein mRNA level [1]. The ribosome conservation equation (v) is equivalent to the previously derived result [3]: $\alpha = (N_{ribo}/P)\beta_r c_p$, where $N_{ribo}$ is the number of ribosomes per cell, $P$ is the total number of amino acids in peptide chains, $\beta_r$ is the fraction of actively translating (bound) ribosomes and $c_p$ is the peptide chain elongation rate.

Explicit expressions for functional and chemical half-lives of bulk protein, and their dependence on the concentration of free ribosomes, can also be derived from Figure 1B, taking into account the negative autoregulation of RNase E (c.f. S2.5 and Eq. S15 in Text S1). For example, one can show that the functional half-life of bulk protein mRNA is given by $T_{1/2,bulk}^{fun} = T_{1/2,bulk}^{fun,o}(1+n_{ribo,free}/L_{m,bulk})$, where $T_{1/2,bulk}^{fun,o}$ is a genetic parameter denoting the functional half-life of bulk mRNA in the absence of ribosomes. Thus, mRNA half-life increases with the probability that the RBS is occupied. This relation reflects translation-degradation coupling trends observed between mRNA degradation and translation [34,39], further discussed in S2.7 of Text S1.

To extract the cell composition from Eq. 2 we require an expression for the gene concentrations, $d_i(\mu)$, of the various gene classes. This expression is given by linking [43] the Cooper-Helmstetter model of DNA replication [77,78] and Donachie's observations regarding the constancy of the initiation volume [41,79]:

$$d_i(\mu) = \frac{1}{V_{ini} \ln 2} \sum_j 2^{-m_{i,j} C \mu} \quad (3)$$

where $V_{ini}$ is the initiation volume, defined as the ratio of the cell volume at the time of replication initiation and the number of origins per cell at that time, $m_{i,j}$ represents the map location of the $j$-th gene in the $i$-th gene class relative to the origin of replication ($0 \le m_{i,j} \le 1$), and finally $C$ is the C period, the time required to replicate the chromosome (roughly 40 min). Recent observations and modeling of the replication initiation mechanism in *E. coli* [41,42] suggest that the initiation volume is regulated to be fixed, and therefore it should be independent of genetic perturbations that do not target that regulation (Tadmor and Tlusty, in preparation). See S2.2 in Text S1 for further details. Thus, we can use Eq. 3 to predict the gene concentration for the genetically perturbed cells considered here.

Equation set 2 provides us with five *equations of state*. We now test whether these equations are consistent with observed WT cell states.

## Results

### The CGGR Model Can Reproduce the WT Cell State

We wish to see whether given measured genetic parameters at a specified growth rate, we can reproduce the cell state, namely the growth rate of the cell and its coarse-grained composition (Table 1).

**Table 2.** CGGR variables, parameters and constants.

| Genetic Parameters | | units |
|---|---|---|
| $V_i^{max}$ | Maximum transcription initiation rate of the i-th gene class | 1/min |
| $U_i^{max}$ | Maximum translation initiation rate of the i-th gene class mRNA | 1/min |
| $K_{m,i}$ | Binding affinity of RNAp holoenzyme to the i-th gene class promoter | $1/(\mu m)^3$ |
| $L_{m,i}$ | Binding affinity of the 30S ribosome subunit to the i-th gene class mRNA RBS | $1/(\mu m)^3$ |
| $t_{c,i}$ | Average time to transcribe the i-th gene class (= $L_i/c_i$) | min |
| $t_{l,i}$ | Average time to translate the mRNA of the i-th gene class (= $L_i/3c_p$) | min |
| $m_{i,j}$ | Map location of the j-th gene in the i-th gene class | dimensionless |
| $T_{1/2,i}^{fun,o}$ | Functional half-life for the i-th gene class mRNA in the absence of ribosomes | min |
| $V_{ini}$ | Initiation volume | $(\mu m)^3$ |
| **Cell state variables** | | |
| $n_{RNAp}$ | Concentration of total RNAp | $1/(\mu m)^3$ |
| $n_{RNAp,free}$ | Concentration of free functional RNAp | $1/(\mu m)^3$ |
| $n_{ribo}$ | Concentration of total ribosomes | $1/(\mu m)^3$ |
| $n_{ribo,free}$ | Concentration of free functional (30S) ribosomes | $1/(\mu m)^3$ |
| $n_{bulk}$ | Concentration of bulk protein | $1/(\mu m)^3$ |
| $\alpha$ | Specific growth rate ($\alpha = \mu \cdot \ln(2)$, where $\mu$ is the doubling rate) | 1/min |
| **Parameters and constants for the unconstrained CGGR** | | |
| $c_{ribo/load}$ | Production cost of one ribosome or load protein | dimensionless |
| $n_0$ | Minimum cell density | $1/(\mu m)^3$ |
| **Parameters and constants for the constrained CGGR** | | |
| $M_{bulk}$ | Bulk protein cutoff | $1/(\mu m)^3$ |
| $c_p^{max}$ | Maximal elongation rate | aa/min |
| $h$ | Hill coefficient | dimensionless |
| **Other parameters, variables and constants** | | |
| $c_p$ | Peptide chain elongation rate | aa/min |
| $c_i$ | RNA chain elongation rate of the i-th gene class | bp/min |
| $C$ | C period | min |
| $d_i$ | Gene concentration of the i-th gene class | $1/(\mu m)^3$ |
| $L_i$ | Length of the i-th gene class | base pairs |
| $v_k$ | Volume of a macromolecule belonging to the k-th species | $(\mu m)^3$ |
| $\Phi$ | Macromolecule volume fraction = $V_{macro}/V_{cell}$ | dimensionless |

Genetic parameters, cell state variables, and other variables and constants associated with the CGGR models. Gene classes labeled by index *i* include: *rrn*, r-protein, bulk protein and any load genes.

For the case of growth at 2.5 doub/h, all genetic parameters, except the Michaelis-Menten parameters for translation initiation ($U_{bulk}^{max}$ and $L_{m,bulk}$) are based on (1) previous estimates derived from empirical data for this growth rate [28], (2) global mRNA half-life measurements at 37°C in LB broth [40], and (3) gene lengths and map locations obtained from the sequenced genome of *E. coli*. These genetic parameters are summarized in Table S1. $U_{bulk}^{max}$ was set at several plausible values (above observed average translation initiation rates [3,80,81] and below the maximum limit where ribosomes are close-packed), with the remaining parameters estimated to minimize the mean square error (MSE) with respect to the WT cell state (Table 1). Errors in estimation of the cell state were no more than 6% of the observed WT cell state and within experimental error bounds of these measurements (~15%; c.f. Table S3 for estimated genetic parameters and corresponding MSEs). Similar results were obtained for the cell state at 1 and 2 doub/h (see for example Table S3 for 1 doub/h). These results indicate that the equations in equation set 2 can be mutually satisfied for these growth rates. We also note that in all cases we found that $L_{m,bulk}$ is of same order of magnitude as the concentration of free ribosome, $n_{ribo,free}$, indicating that the RBSs are not saturated by free ribosomes, in agreement with pervious studies [30–33]. Further details are given in S1.1.1 of Text S1.

### rRNA Operon Inactivation Experiments: The Squires Data

In the series of experiments that we consider here, Asai et al. [19] have measured the growth rate and rRNA to total protein ratio of seven *E. coli* strains, with rRNA operon copy numbers ranging from one to seven per chromosome (Figure 2). Since all strains were grown in a constant environment of Luria-Bertani broth at 37°C ($\mu = 2.0$ doub/h for the WT strain), the CGGR model is applicable. We first reconstructed the WT genetic parameters and the relevant physical constants (C periods and elongation rates) for a growth rate of 2 doub/h (c.f. Table S5 and S1.2 in Text S1 for a detailed account). Next, by analyzing the published lineage of these strains (Table S6) we derived the genetic parameters for each specific strain (Table S7). In S1.3 of Text S1 we explain which genetic parameters for the WT cells can be carried over to the inactivation strains and which parameters change, and how. The WT cell state at 2 doub/h is given in Table S2 and the genetic parameters at 2 doub/h for the WT cell and the inactivation strains are summarized in Table S5 and Table S7, respectively.

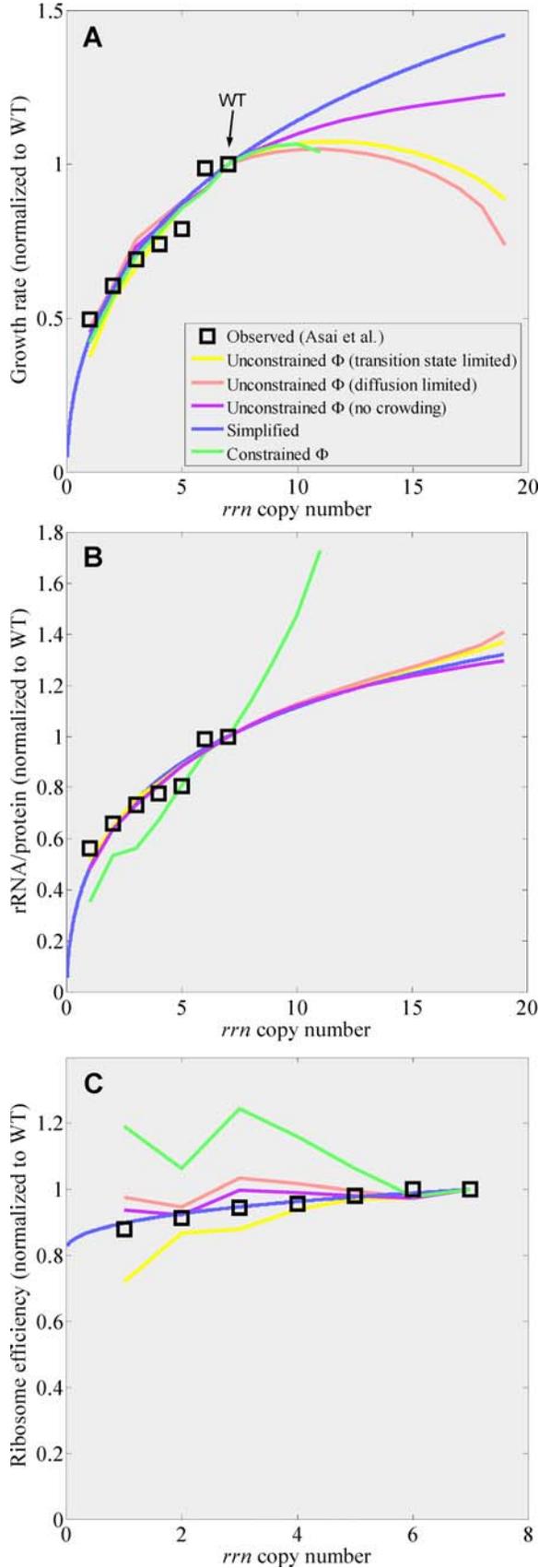

**Figure 2. Comparison of rRNA operon inactivation data of Asai et al. [19] to CGGR models and predictions for higher rRNA operon copy numbers.** (A) Growth rate as a function of rRNA operon copy number per chromosome. The maximum standard error of growth rate measurements was 0.07 [19]. (B) rRNA to total protein ratio, where total protein is given by total amino acids in the form of r-proteins and bulk proteins. Measurement error was not available for this data. In the case of the constrained model, solutions were not obtainable above a copy number 11. (C) Ribosome efficiency, defined as $e_r = \alpha \cdot P/N_{ribo}$ [3,19] (see text), was obtained by dividing the growth rate in (A) by the ratio of rRNA to total protein in (B). All curves are normalized to WT values at 2.0 doub/h. The legend to all figures is given in (A). The kink observed for copy number 2 is due to strong expression of lacZ in this strain used for inactivation. Beyond the WT rRNA gene dosage, rRNA operons were added at the origin (also see S1.4 in Text S1). The rRNA chain elongation rate, $c_{rrn}$, was assumed to be fixed in these simulations.

Given the genetic parameters, we set to solve Eq. 2 for the different strains. However, in order to solve for the CGGR cell state, which consists of six state variables, we need an additional relation which apparently does not arise from kinetic considerations. A hint to the solution may lie in the fact that so far we have neglected the *function* of the bulk protein and biophysical considerations such as macromolecular crowding.

## Macromolecular Crowding and the Function of the Bulk Protein

The *in vivo* milieu of *E. coli* is extremely crowded with macromolecules [45] with typical values of macromolecule volume fraction $\Phi = V_{macro}/V_{cell}$ of 0.3–0.4 [46]. Observations of WT *E. coli* in varying environments suggest that the macromolecular mass density of the interior of the cell is more or less a constant [23]. If we neglect the contribution of RNAp, mRNA and DNA (~6% at 2.5 doub/h [17]) this is roughly equivalent to stating that

$$\Phi \equiv v_{ribo} n_{ribo} + v_{bulk} n_{bulk} + v_{load} n_{load} = \text{const} \qquad (4)$$

where $v_i$ is the volume occupied by a particle belonging to the $i$-th species (c.f. Table S2), and with potential contribution from "load genes" that express products not utilized by the cell and pose a pure burden, like antibiotic resistance for example. Equation 4, which balances bulk protein *against* ribosomes, leads to a contradiction: it appears from this model, that by genetic perturbations, e.g. by increasing the rRNA operon copy number, one could construct a cell composed almost entirely of ribosomes with no bulk proteins to support it, or vice versa. To resolve this difficulty we need to take into account the fact that some of the bulk proteins are required to support ribosome synthesis.

One possible resolution is to introduce a mechanism that would limit protein and ribosome synthesis when bulk protein density is reduced. For example, one could assume that the peptide chain elongation rate, $c_p$, is given by $c_p = c_p^{\max} / \left[1 + (M_{bulk}/n_{bulk})^h\right]$, where $h$ is some Hill coefficient, $c_p^{\max}$ is the maximal elongation rate and $M_{bulk}$ is a cutoff. $M_{bulk}$ may depend on the environment, reflecting the dependence of $c_p$ on the environment [17]. Note that $c_p$ affects our system of equations through the translation times $t_{l,i}$.

This criterion along with Eq. 2 and Eq. 4 define the *constrained CGGR* model.

However, is the macromolecular volume fraction, $\Phi$, really constant? The phenomenological evidence indicating that $\Phi$ is roughly constant has been obtained for WT cells in different environments, and not for a suboptimal mutant growing in a given environment like the Squires strains. Indeed, it has been proposed that $\Phi$ can vary by adjusting the level of cytoplasmic water to counter changes in the external osmotic pressure [82]. These observations suggest that $\Phi = $ const is apparently not a universal law in *E. coli*.

An alternative resolution, which does not hypothesize that $\Phi = $ const, could be to postulate a *cost criterion*, which states that the amount of ribosomes that the cell can produce is limited by resources, such as ATP, amino acids, etc., that are made available by the bulk proteins. Assuming that bulk protein concentration, $n_{bulk}$, is proportional to its demand, i.e. to total ribosome concentration, $n_{ribo}$, and also to possible load protein concentration, $n_{load}$, the criterion takes the form:

$$n_{bulk} = n_0 + c_{ribo} n_{ribo} + c_{load} n_{load} \quad (5)$$

where $c_i$ are the costs and $n_0$ is some minimal density of the cell (e.g. housekeeping proteins, membrane building proteins etc.), assumed to be more or less constant. $c_{ribo}$, for example, is defined as the number of bulk proteins per cell, $N_{bulk}$, required to increase the number ribosomes in the cell, $N_{ribo}$, by one, given a fixed environment $E$ (i.e. sugar level, temperature, etc.), a fixed cell volume and a fixed number proteins, $N_j$, expressed from all other genes (akin to the definition of a chemical potential):

$$c_i = \left(\frac{\partial N_{bulk}}{\partial N_i}\right)_{E, V_{cell}, N_{j \neq i}} \quad (6)$$

In other words, to synthesize and support one additional ribosome per cell, in a constant environment, cell volume etc., according to this definition, would require an additional $c_{ribo}$ bulk proteins per cell ($c_{ribo}$ is dimensionless). An equivalent way to interpret Eq. 5 is to say that $c_{ribo}$ is the *capacity* of a ribosome to synthesize bulk proteins: one additional ribosome added to the cell will synthesize $c_{ribo}$ bulk proteins. Thus, at steady-state, cost and capacity are different sides of the same coin. The costs, $c_i$, depend on the environment since the cost of producing and maintaining a ribosome in a rich environment is expected to be lower than the cost in a poor environment due to the availability of readymade resources that otherwise the cell would need to produce on its own. The hypothesized costs, $c_i$, can therefore be thought of as *effective* environment-dependent genetic parameters and could, in principle, be estimated from knowledge of the genetic networks invoked in a given growth environment. Note that Eq. 4 is actually a special case of Eq. 5 for certain negative costs. Eq. 5 also crystallizes the difference between bulk proteins and load proteins: the latter are a burden for the former. The cost criterion together with Eq. 2 define the *unconstrained CGGR* model. The final equation set for both models is summarized in S2.6 of Text S1.

From an experimental point of view it should be possible to discern between the two hypotheses: one model (Eq. 5) predicts a positive slope for the $n_{bulk}$ vs. $n_{ribo}$ curve, whereas the other model (Eq. 4) predicts a negative slope. In the Discussion we suggest how the cost criterion may naturally occur in the cell.

**Global Crowding Effects.** Since macromolecular volume fraction can change in the unconstrained CGGR model due to Eq. 5, it is essential to examine how crowding can affect the input genetic parameters. We considered two possible crowding scenarios (c.f. S2.4 in Text S1). In the "transition state" scenario it was assumed that holoenzyme-promoter and 30S-RBS binding affinity are transition state limited, that is, the probability that an association complex will decay to a product is small compared with the probability that it will dissociate back into the reactants [83]. Typically such reactions display an increase in efficiency as crowding is initially increased and eventually decrease in efficiency since in the limit of high fractional volume occupancy, all association reactions are expected to be diffusion limited and hence slowed down [44]. The forward rate of transition state reactions is predicted to display a bi-modal dependence on the macromolecular volume fraction $\Phi$ [44,51,83–85]. Such a bi-modal dependence has been observed experimentally *in vitro* [44,86]. Assuming binding affinities weaken in the limit of high volume fraction we expect that in such a case the binding affinity will display a maximum (Figure S8A). In the transition state binding scenario we have further assumed the binding affinity has been evolutionary tuned to be maximal at the WT value of $\Phi$ ($= 0.34$ [46]), similar to the temperature optimum commonly exhibited for RNAp/promoter complexes [75].

A second, "diffusion limited" scenario, assumed that all reactions were diffusion limited, that is almost every association complex will become a product [83]. For this scenario, binding affinities were assumed to decay exponentially (Figure S8B), as has been observed *in vitro* for diffusion coefficients [47,51,87], and suggested for the forward rate in diffusion limited reactions [44,51,83–85]. Thus, both models assumed that binding affinities decay at high $\Phi$ ($>0.34$), mainly due to diffusion limited forward rates [44,50,83]. Surprisingly, we found that predictions were quite insensitive to the exact crowding scenario implemented, due to a homeostasis mechanism that arises from internal feedbacks and compensates for moderate crowding effects ($\Phi < 0.3$–$0.4$). This effect is further discussed below.

One may suspect that mRNA degradation would also be influenced by crowding because it involves the association of two macromolecules (Figure 1B). Yet interestingly, due to the negative autoregulation of RNase E [57,61,88], which adjusts its steady-state expression to that of its substrates [61], the crowding effects on the binding affinity of RNase E to the 5′ end of the mRNA appear to cancel out with the crowding effects on the binding affinity of RNase E to its own mRNA (c.f. S2.7 in Text S1 for more details).

### Comparison to the Experimental Data of Squires

With the CGGR models at hand, we now compute the cell states for each of the seven strains used in the Squires rRNA operon inactivation experiments. We will use this data to fit for the unknown environment dependent parameter in each of the CGGR models: $c_{ribo}$ for the unconstrained model and $M_{bulk}$ for the constrained model. In the case of the unconstrained model, the predicted rRNA to total protein ratio was more sensitive to $c_{ribo}$ than the predicted growth rate, with a best fit for the former at $c_{ribo} \approx 38$ bulk proteins per ribosome (for mean square errors refer to Figure S1). For comparison, a 70S ribosome is about 70 bulk proteins in mass. Note that $c_{ribo}$ has a rather limited range of values since $0 < c_{ribo} < n_{bulk}/n_{ribo} \simeq 101$ via Eq. 5.

For the constrained CGGR model, the minimum Hill coefficient to yield a solution that did not diverge in growth rate for copy numbers greater than 7, which contradicts observation (see Introduction), was $h = 2$ (see e.g. Figure S2 for a fit with $h = 1$). For $h = 2$, $M_{bulk}$ was chosen minimize the MSE with respect to the growth rate, which displayed a minimum, and the best fit was achieved for $M_{bulk} \cong 7.4 \cdot 10^6$ molec/WT cell (for all MSEs c.f. Figure S1). Attempting to minimize the MSE with respect to the rRNA to total protein ratio resulted in a slightly lower MSE (though still higher than the MSE for the unconstrained model fit), however solutions diverged in growth rate for copy numbers greater than 7, again contradicting observation. Increasing the Hill coefficient so as to penalize the peptide chain elongation rate, $c_p$, for higher copy numbers did not remedy this and growth rate continued to diverge for copy numbers greater than 7 (c.f. Figure S3) rendering such solutions inapplicable. Finally, increasing the Hill coefficient beyond 2 did not improve the overall MSE to either the growth rate or to the rRNA to total protein ratio (Figure S1). Thus the fit for the constrained model presented here represents the best fit, over all parameter range, which does not contradict observation.

Figures 2A and 2B show the observed growth rate and rRNA to total protein ratio plotted against the best fits of these models. In both cases, the fit to the observed data was reasonable, however the model for which macromolecular volume fraction, Φ, was not constrained gave an overall better fit indicating a preference for that model. Further evidence in favor of this model and against the constrained model will be considered in the Discussion. The deviation observed for the Δ6 strain may possibly be due to tRNA imbalance in this strain [19].

## Free RNAp and Free Ribosomes Self-Adjust to Counter Changes in Binding Affinities Due to Crowding

Whereas the macromolecular volume fraction Φ in the constrained model is, by definition, constant, the unconstrained CGGR model predicts that Φ increases with the number of rRNA operons with consequences on binding affinities (Figure 3). This increase in the macromolecular volume fraction is due to an increase in both ribosome concentration and bulk protein concentration due to the relation imposed by the cost criterion (Eq. 5; also c.f. Figure S5). Quite surprisingly, the fit to the Squires data depends very little on the crowding scenario chosen. This results from a self-adjusting homeostasis mechanism: it is the ratios of free RNAp and free ribosomes with respect to their corresponding binding affinities that govern the transcription and translation rates (Eq. 1). Hence, although the binding affinities change with Φ, the concentrations of free RNAp and free ribosomes counterchange to stabilize these ratios (see Figure S4 and S1.6 in Text S1). The efficiency of the homeostatic mechanism diminishes as the degree of crowding is increased above ∼0.4, as can be seen by comparing to predictions of the "no crowding" scenario, in which binding affinities were assumed to be independent of Φ (Figure 2A and 2B).

## Translation-Degradation Coupling

Due to translation-degradation coupling, bulk mRNA half-life was predicted to mildly increase with rRNA operon copy number for all models. In both crowding scenarios, bulk mRNA half-life increased from about 0.8 of the WT half-life to about 1.2 of the WT half-life. The increase in mRNA half-life is caused by the increase in the ratio of the RBS binding affinity and the concentration of free ribosomes with rRNA operon copy number (Figure S4). This ratio reflects the probability that a RBS is occupied, thereby protecting the mRNA from cleavage.

## Beyond a Copy Number of Seven

Increasing the rRNA operon copy number beyond 7 (at map location 0), we found that both CGGR models exhibit a shallow optimum plateau for of the growth rate in the range of 7–12 copies, with the maximum occurring at a copy number of 10–11 for the diffusion limited scenario, and 11–12 for the transition state scenario. In the case of the unconstrained models, overcrowding contributed to the formation of this maximum (e.g. there is no maximum in the unrealistic model where binding affinities are assumed to be independent of Φ). A striking difference between the models is in their predictions regarding the rRNA to total protein ratio. This ratio strongly diverges in the constrained model at high copy numbers because ribosomes are formed at the expense of bulk protein (see Discussion).

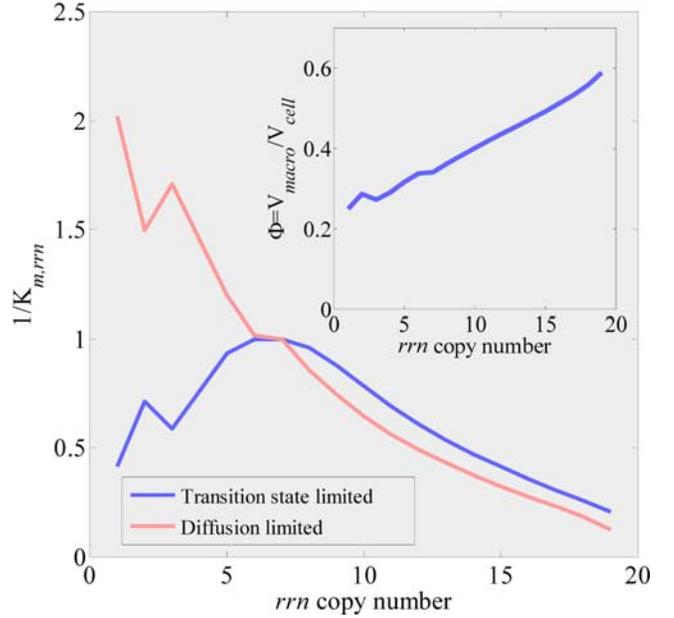

**Figure 3. Predicted effect of crowding on the rRNA promoter binding affinity for two crowding scenarios.** In the transition state scenario, binding affinities initially strengthen as macromolecular crowding is increased due to increased entropic forces, while in the diffusion limited scenario binding affinities weaken as macromolecular crowding is increased due to increased diffusion times. In both cases, binding affinities weaken when macromolecular crowding is increased beyond the WT crowding state due to increased diffusion times (see also Figures S8 and S2.4 in Text S1). Quantitatively, affinities can vary by up to a factor of 5 (transition state limited scenario) to 16 (diffusion limited scenario), measured as the ratio of the maximum and minimum values of binding affinities in the range of rRNA operon copy numbers considered. Similar results are obtained for RBS binding affinities. Binding affinities were normalized to WT values. Insert: predicted macromolecular volume fraction, Φ, as a function of rRNA operon copy number per chromosome.

## A Simplified Model

For the data of the Squires strains, the unconstrained CGGR can be approximated by a simplified three-state model involving only $n_{ribo}$, $n_{bulk}$, and μ (c.f. S3 in Text S1):

$$\begin{aligned}&(i)\ n_{ribo} = g_{rrn}/\mu\\&(ii)\ n_{bulk} = g_{bulk}n_{ribo}/\mu\\&(iii)\ n_{bulk} = n_0 + c_{ribo}n_{ribo}\end{aligned} \quad (7)$$

where $g_{rrn}$ and $g_{bulk}$ are effective genetic parameters that are estimated from the WT cell state. Equation (i) reflects ribosome synthesis, (ii) reflects bulk protein synthesis and (iii) is the cost criterion. Interestingly, in a different context of WT cells measured in varying environments, a relation similar to Eq. (ii) has been observed [21]. Solving Eq. 7 for the growth rate we obtain

$$\mu = \frac{c_{ribo}g_{rrn}}{2n_0}\left(\sqrt{1+4\frac{n_0 g_{bulk}}{g_{rrn}c_{ribo}^2}}-1\right) \quad (8)$$

In the limit $g_{rrn}\to\infty$, $\mu \leq g_{bulk}/c_{ribo}$, suggesting that in the absence of crowding effects, growth rate would be limited by the production cost of a ribosome. To fit to experiments where the rRNA operon copy number is manipulated, we approximate that $g_{rrn}\to g_{rrn}\cdot$copy #/7. The best fit to the Squires data was obtained for $c_{ribo}\simeq 38.2\pm 2.8$, in agreement with the prediction of the full unconstrained model (see Figure 2A and 2B, and also Figure S1 for MSEs). Since the simplified model is unrealistic in the sense that it lacks crowding effects, growth rate continues to increase with rRNA operon copy number.

### Ribosome Efficiency

Ribosome efficiency has been previously defined as $e_r \equiv \beta_r c_p = \alpha P/N_{ribo}$ [3,19] where $c_p$ is the peptide chain elongation rate and $\beta_r$ is the fraction of actively translating ribosomes. For wild-type cells, $\beta_r = 80\%$, and is independent of growth rate [17]. Genetically perturbed cells may however respond differently [19]. For example, the simplified model predicts that $e_r = \alpha P/N_{ribo} = \ln 2(g_{bulk}L_{bulk}+\mu L_{r\text{-}protein})$, where $L_i$ is the length of gene class $i$ and $\mu$ is given by Eq. 8. Since $c_p$ is assumed to be fixed in the unconstrained/simplified models, ribosome efficiency is therefore expected to decrease purely due to kinetic considerations. Crowding effects tend to either increase or decrease ribosome efficiency, depending on the scenario. In Figure 2C we plot the ribosome efficiency for the various crowding scenarios in the unconstrained CGGR model, for the constrained CGGR model and for the simplified three-state model. We see that the data points lie between the diffusion limited crowding scenario and the transition state crowding scenario, which possibly indicates that the *in vivo* crowding scenario is somewhere between being diffusion limited and transition state limited. Overall however, the diffusion limited model was a better predictor of ribosome efficiency than the transition state model and its deviation from the observed data points was on the order of the maximum error for these points (the maximum deviation from experimental data points is ~10%, and although the error for the protein measurement was not stated in [19], the maximum error on ribosome efficiency was at least 9% based on the errors quoted in [19]). This result possibly indicates a preference for the diffusion limited scenario for the *in vivo* case (see Discussion). The solution for which binding affinities are independent of crowding (the "no crowding" scenario) also fits the data due to the proposed homoeostasis mechanism for $\Phi<\sim 0.4$. The constrained model clearly deviates from the experimental points indicating, as we have seen before, that the constrained CGGR model is not applicable to *E. coli*. Finally, the simplified model appears to adequately trace the observed ribosome efficiency.

### The Initiation Rate of a Single rRNA Operon

Figure 4 shows the initiation rate of a single rRNA operon, $V_{rrn}$, as a function of the rRNA operon copy number as predicted by the unconstrained model (Eq. 1). The solid lines represent models where the rRNA chain elongation rate was assumed to be constant (85 nuc/sec [17]; Table S5). Both unconstrained models exhibit an increase in rRNA expression per operon as copy number is decreased from 19 copies per chromosome down to 3 copies per chromosome (in the case of the diffusion limited model) and 5 copies per chromosome (in the case of the transition state model). This trend is in agreement with the feedback response mechanism, especially for the diffusion limited model (see Discussion). It has been shown that the rRNA chain elongation rate (but apparently not mRNA chain elongation rate) increases in inactivation strains from ~90 nuc/sec in a WT strain to ~135 nuc/sec in a strain with four inactivated rRNA operons [54], but remains constant in strains with increased rRNA gene dosage [65]. To check how these finding affect our predictions, we also included a model where rRNA chain elongation rate decreased linearly from 160 nuc/sec for one functional rRNA operon per chromosome, to 85 nuc/sec for the WT strain (dashed lines in Figure 4). Indeed, the feedback response seems to be stronger for the inactivation strains when assuming that rRNA chain elongation rate increases as more operons are inactivated. Quantitatively, for the diffusion limited model, rRNA expression from a single operon increased from 0.6 of the WT expression for 19 chromosomal rRNA operons to about 1.1 of the WT expression for 3 chromosomal rRNA operons. Finally, the "no crowding" scenario exhibited a milder feedback response due to departure from the homeostasis discussed earlier.

### Discussion

The goal of the coarse-grain genetic reactor (CGGR) approach is to attempt to link global phenotypes, such as growth rate and cell composition, directly to genetic parameters, while keeping the model as simple as possible by means of coarse-graining. The present CGGR models assumed the simplest type of environment, namely a spatially and temporally constant environment that is unlimited in resources. The models attempt to explain a series of experiments performed by the Squires group [19] in which growth rate and cell composition have been measured for seven *E. coli* strains with varying rRNA operon copy numbers. The genome of all seven strains has been coarse-grained, and their corresponding cell state was calculated based on the CGGR models.

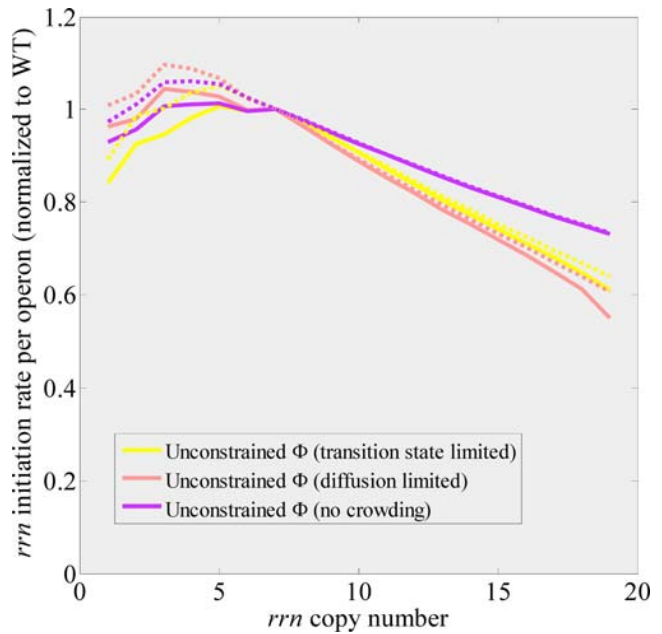

**Figure 4. Initiation rate of a single rRNA operon ($V_{rrn}$) for various crowding scenarios in the unconstrained CGGR model.** Solid lines are predictions for the case where the rRNA chain elongation rate is assumed to be fixed at its WT value of 85 nuc/sec [17], while the dashed lines take into account the observed effect of rRNA operon copy number on rRNA elongation rates [54,65] (see text for further details). $V_{rrn}$ is given by Eq. 1. In each case considered, $c_{ribo}$ was obtained by fitting to the Squires data.

We considered two possible CGGR models, one in which the macromolecular volume fraction is constrained to be fixed, and one in which macromolecular volume fraction is unconstrained. We have seen that the unconstrained CGGR model appears to give an adequate fit to experimental data, while the fit for the constrained CGGR model is rather poor (despite the latter having an additional degree of freedom). Yet beyond the fit of the unconstrained model to the Squires data, this model also appears to be consistent with additional observations regarding strains with altered rRNA operon copy numbers. For example, the unconstrained CGGR model predicts that growth rate decreases for higher *rrn* copy numbers, as indicated by observation. For comparison, the best fit of the constrained CGGR model actually predicted that growth rate increases for rRNA operon copy numbers greater than 7, contradicting observation. In addition, both models predict that the concentration of ribosomes (and ribosomes per cell) decreases with rRNA operon copy number (Figure S5), as was shown in measurements of an earlier set of inactivation strains engineered by the same group, with rRNA operon copy number ranging from three to seven [54]. Below we discuss further evidence in support of the unconstrained CGGR model: observations regarding the nucleoid size in strains with altered rRNA gene dosage appear to be consistent with the crowding predictions of this model. Finally, the unconstrained (diffusion limited) CGGR model is in agreement with the trend associated with the feedback response, and appears to be in qualitative agreement with measurements of this effect. The proposed model is also consistent with the feedback model proposed by Nomura and coworkers, as will be discussed further below. The constrained CGGR model, on the other hand, in addition to yielding an inferior fit to the Squires data, is also problematic from a biological standpoint. This model should predict that ppGpp levels rise due to a shortage in an essential factor such as charged tRNAs [3,6,89]. However, ppGpp levels were observed to be constant in similar rRNA inactivation strains with up to four inactivations [54]. In addition, the constrained model appears to be considerably more complicated than the unconstrained model in that it necessitates some kind of homeostasis mechanism for keeping the volume fraction fixed, to which there is no experimental evidence as far as we know, while the unconstrained model does not necessitate any additional biological mechanisms (see below). In fact, evidence from osmotically stressed cells indicates that the volume fraction of macromolecules can change quite considerably [82]. Indeed, these experiments indicate growth rate can be limited by crowding [82], just as predicted by the unconstrained CGGR model (see below).

Since the macromolecular volume fraction in the unconstrained CGGR model is not constant, we needed to consider crowding effects on association reactions such as transcription initiation and translation initiation. We investigated two possible crowding scenarios: one in which all association reactions are diffusion limited and one in which all association reactions are transition state limited and have been evolutionarily tuned to be maximal at the WT volume fraction. Both crowding scenarios give an adequate fit to the growth rate and rRNA to total protein ratio data, thanks to the homeostasis mechanism involving free RNAp and free ribosomes. However, the diffusion limited model seems to give a slightly better fit when considering the feedback response and ribosome efficiency data, possibly indicating a preference for this model. Indeed, it has been proposed that the *in vitro* 30S-mRNA association may be diffusion rate limited since *in vitro* measured on rates are of the order of the diffusion limit [90]. In addition we have proposed a simplified version of the unconstrained CGGR model, which is a three-variable model and is included since it is an analytically solvable reduction of the more complicated six state model. We have shown, however, that since the simplified model does not take into account the physical effects of crowding, its predictions for strains with increased rRNA operon gene dosage is unrealistic. Hence the full unconstrained CGGR model is the biophysical model that we propose to be relevant for *E. coli* growing in balanced exponential steady-state growth in a rich medium.

### Nucleoid Compaction in the Inactivation Strains

Further support for the reduction of macromolecular volume fraction in the rRNA inactivation strains may perhaps be found in fluorescence images of the WT Squires strain vs. the Δ6 strain in which six rRNA operons have been inactivated (Figure 4 in [19]). The nucleoid in the WT cells is seen to be much more compact than in the Δ6 strain, suggestive of lower entropic forces in the latter strain due to a lower degree of crowding [48]. Recent observations in strains in which six rRNA operons were entirely deleted from the genome (and not just inactivated as in [19]) show similar results, and also indicate that the compact structure of the nucleoid was recovered in strains in which rRNA is expressed solely from a high copy number plasmid with all other *rrn* operons entirely deleted from the genome (S. Quan and C. L Squires, personal communication). These results are consistent with crowding effects [48] predicted by the unconstrained CGGR model, effects that are absent in the constrained CGGR model.

### The Feedback Response of the rRNA Operons

While both unconstrained CGGR models exhibited a decrease in the expression of a single rRNA operon as rRNA gene dosage was increased, as is observed in the feedback response, in the case of the inactivation strains, the diffusion limited model appeared to be in better agreement with the feedback response than the transition state model (Figure 4). In the former model, rRNA expression from a single rRNA operon increased as rRNA operon copy number was decreased from 19 copies per chromosome to 3 copies per chromosome. The transition state model exhibited this dependence only up to an rRNA operon copy number of 5. The increase in the rRNA operon expression is due to an increase in the ratio of free RNAp concentration and the rRNA operon binding affinity (Figure S4). It is interesting to note that in the diffusion limited scenario, it is actually the changes in binding affinities, and not free RNAp, which correct for the observed trend of the feedback response, as free RNAp concentration is actually predicted to increase when the number or rRNA operons per chromosome is increased (Figure S4B). Furthermore, we found that a model in which the rRNA chain elongation rate increases when inactivating rRNA operons, as observed experimentally [54], exhibits a slightly stronger feedback response when inactivating rRNA operons than a model that assumes that this parameter is constant.

Quantitatively, in the case of the diffusion limited scenario with variable rRNA chain elongation rate, the rRNA expression from a single operon increased from 0.6 of the WT expression for 19 chromosomal rRNA operons to about 1.1 of the WT expression for 3 chromosomal rRNA operons. Although rRNA operon synthesis rate was not measured for the inactivation strains considered here, we can qualitatively compare these predictions to experiments with other strains. Strains in which four rRNA operons were inactivated exhibited a 1.4 to 1.5 increase rRNA operon expression relative to a WT background, where expression was measured as $\beta$-galactosidase activity from WT P1 promoter fragments fused to a *lacZ* reporter gene and normalized to expression from a WT background [64] (we are not aware of measurements for lower copy numbers). In a similar manner rRNA expression was shown to decrease by a factor of 0.65 to 0.8 with respect to the WT background in strains in which rRNA gene dosage increased by using plasmids expressing rRNA (the plasmid copy number was not specified) [64]. In a different study by the same group, the initiation rate in strains with increased rRNA operon copy number was obtained based on counting the number of RNAp bound to rRNA operons using electron microscopy and measurement of the rRNA elongation rate, and yielded 0.66 of the WT initiation rate [65]. Although the predicted feedback response for the inactivated strains is somewhat weaker than the response observed experimentally, the overall trend appears to be in qualitative agreement with the feedback response, i.e. as the rRNA operon copy number is increased, the transcription from a single rRNA operon decreases. We note however that the genetic makeup of the inactivated strains tested above differed from the inactivated strains of Asai et al. [91], especially in the respect that in the former strains, each inactivated rRNA operon expressed antibiotic resistance, which may have had adverse effects on the cell. The fact that the elicited feedback response is not as strong as the one observed experimentally in the inactivation strains may also possibly be a consequence of some of the simplifying assumptions made in this model (e.g. ideal RNAp autoregulation or the somewhat naïve crowding models assumed) or perhaps indicate the presence of an additional mediator (see below).

The unconstrained CGGR model also predicts that bulk mRNA transcription would be affected by the change in rRNA gene dosage since in the current model bulk RNAp binding affinity has the same response to changes in macromolecular crowding as the rRNA binding affinity. The effect may be, however, somewhat alleviated by the fact that bulk mRNA binding affinity is proposed to be about 3 times stronger than the P1 rRNA promoter (which is the major site for the feedback response [92]) at this growth rate (Table S4), thus closer to saturation, and can even be ~30 times stronger in poor medium (Table 2 in [28]), although is has also been suggested that RNAp promoters may require the same or less RNAp than other RNA promoters for transcription [93]. Also, in principle rRNA and bulk promoters could respond differently to crowding. When measured experimentally, mRNA promoters did in fact exhibit some reduction when the feedback response was induced using increased rRNA gene dosage: while expression of a P1-*lacZ* fusion decreased by 0.45 relative to a control with WT rRNA gene dosage, *spc* or *lac*UV5 promoters fused to *lacZ* decreased by ~0.8 relative to the same control [92]. Nevertheless, these results may also indicate that there is an additional mediator involved, which interacts specifically with the P1 rRNA promoter [92]. If this turns out to be the case, the influence of such an effector could be incorporated into the proposed model.

No molecule, however, has yet been implicated in the feedback response to a change in the rRNA gene dosage. In addition, experiments indicate that ribosomes appear not to be directly responsible for this feedback response (see Introduction). Therefore, it may be possible that for the type of perturbation considered here, the feedback response results, at least in part, from internal feedbacks inherent in the system. Various models have suggested that free RNAp is in one way or another limiting (e.g. [28] and also discussion in [3,5]), yet it is not certain that changes in RNAp alone can account for the observed changes in rRNA expression due to changed rRNA gene dosage [1,5,65]. In the present work, we are only concerned with the response of the cell to changed rRNA operon copy number in a constant rich environment, where ppGpp concentration is constant. Therefore we do not attempt to explain how ppGpp modulates rRNA expression. In addition, we found that the model that best fits experimental data is one where both the concentration of free RNAp and the binding affinities of RNAp to its promoters are altered in response to changes in rRNA gene dosage. Therefore, according to this model, it is not the concentration of free RNAp which affects the transcription, as has been proposed in the past, but rather the ratio of the concentration of free RNAp to its binding affinity that determines transcription. In fact, we have seen that in the diffusion limited scenario, free RNAp concentration actually decreases as rRNA operon copy number is reduced, and it is the increase in the rRNA operon binding affinity that is responsible for the increased transcription of the rRNA operon (e.g. Figure S4B).

The notion that crowding can be an effector modulating transcription of the rRNA operons is consistent with the feedback model of Nomura and coworkers [1,63] since only functional rRNA gets assembled into ribosomes, and together with supporting bulk proteins crowd the cell, thus contributing to the feedback response. Nonfunctional rRNA would be degraded away and hardly contribute to crowding or the feedback response. Finally, the notion that the feedback arises from the inherent internal feedbacks in the cell is consistent with the indirect aspect of the feedback response proposed by Nomura and coworkers [62].

### Effect of Increased rRNA Operon Copy Number on Growth Rate

Extrapolating to higher copy numbers suggests that the WT growth rate in a constant and rich environment is nearly maximal. In an experiment with increased rRNA gene dosage, where ppGpp concentration was shown to be constant, the growth rate of a strain carrying extra rRNA operons on a plasmid indeed decreased by 22% relative to a WT strain carrying a control plasmid expressing nonfunctional rRNA [20], in agreement with the trend predicted by the model. In another experiment with increased rRNA gene dosage, growth rate decreased relative to WT cells containing a control plasmid, and rRNA to total protein ratio was more or less constant (thus appearing to favor the unconstrained CGGR model) although the authors argue that there may be tRNA imbalance in these strains [94]. In addition, the unconstrained CGGR model predicted that ribosome and bulk protein concentration increase with rRNA operon copy number (Figure S5) thus leading to an increase in the macromolecular volume fraction (Figure 3, insert). This increase is due to the cost criterion hypothesis (Eq. 5), which correlated the concentration of bulk protein in the cell with the concentration of ribosomes.

### The Optimum in Growth Rate

The biophysical origin of the predicted upper limit on growth rate with respect to the rRNA operon copy number, suggested by the unconstrained CGGR model, is overcrowding of the cytoplasm with ribosomes and with bulk proteins supporting/ synthesized by those additional ribosomes via the cost criterion relation (Eq. 5). As rRNA operon copy number is increased, the concentration of ribosomes and bulk protein increases (Figure S5) leading to an increase in macromolecular volume fraction in the cell (Figure 3, insert). *In vitro* experiments suggest that in a crowded environment diffusion times increase [47,51,87]. If in an overcrowded environment, when all reactions are thought to be diffusion limited [44,51,83–85], increased diffusion times cause binding affinities to weaken, then overcrowding will reduce the efficiency of transcription initiation and translation initiation (Figure 4 and Figure S4). This reduction in efficiency ultimately causes the growth rate to decrease at high rRNA operon copy numbers. In the scenario where binding affinities were assumed to be independent of the level of crowding in the cell (the 'no crowding' scenario in Figure 2A), growth rate continued to increase as rRNA operon copy number increased, indicating that the reduction in growth rate in the transition state and diffusion limited crowding scenarios was due to crowding effects. See also Figure S6 for a breakdown of the different contributions in the ribosome synthesis equation, Eq. 2iii. Interestingly, a similar phenomenon may be occurring in osmotically stressed cells. It has been shown experimentally that the growth rate of osmotically stressed cells is correlated with the amount of cytoplasmic water in those cells [82] leading those authors to propose that increased diffusion times of biopolymers due to crowding may be limiting growth rate. This conclusion appears to be in accord with our findings.

The fact that the maximum in growth rate is so shallow may suggest that in a natural environment for *E. coli* there are additional constraints in the system. In nature, *E. coli* is likely to experience chronic starvation conditions like in water systems, as well as fluctuating environments like in the host intestine [15,16]. Indeed, it has been shown that *E. coli*'s growth rate displays a more pronounced dependence on the rRNA operon copy number in a changing environment compared to a constant one [15], and that a high rRNA operon copy number enables *E. coli* and other bacteria to adapt more quickly to changing environments [15,95,96].

Finally we wish to point out that the optimum we have shown is only with respect to *rrn* copy number perturbations of a WT *E. coli* genome, and therefore may possibly not be a global one. A higher growth rate could perhaps be attained when considering perturbations of all genetic parameters.

## Efficiency and Decoupling of the Replication Initiation Module

The unconstrained CGGR model suggests possible insights into the design principles of *E. coli*. The model introduces the concept of a cost per gene class, akin to a chemical potential. In the absence of load genes for example, the *cost criterion* basically measures the number of bulk proteins needed to support the synthesis of ribosomes (or vice versa). This criterion implies that *the cell is efficient*: bulk protein is utilized to its full potential and is not stored as inventory for later use. This is true even for genetically perturbed (i.e. suboptimal) cells. A similar notion of efficiency was suggested by Ecker and Schaechter in the context of WT cells growing in different environments [21]. How then is the cost criterion realized by the cell? Perhaps the cost criterion is realized simply by virtue of internal feedback. If, for example, the rRNA operon copy number is slightly increased, resulting in a small increase in ribosome concentration, $\Delta n_{ribo}$, the transient deficit in bulk protein ($-\Delta n_{bulk}$) will be compensated for, at steady-state, by the extra ribosomes when $\Delta n_{bulk}$ ($=c_{ribo}\Delta n_{ribo}$) bulk proteins are synthesized. $n_{bulk}$ therefore increases to the *minimum* concentration needed to sustain these excess ribosomes. Thus, the cost criterion obviates the need for a homeostatic mechanism for keeping $\Phi$ fixed. Nevertheless, direct experimental proof for the cost criterion is currently lacking.

An additional engineering principle suggested by the CGGR models is related to the DNA replication mechanism. Replication enters the model through the C period and the initiation volume (Eq. 3), both of which are regulated to be roughly constant [23,41] and thus in principle unaffected by genetic perturbations (Tadmor and Tlusty, in preparation). Since this implies that gene concentrations do not depend strongly on growth rate (see Figure S7 and S2.2 in Text S1), this result suggests that the regulatory mechanism of replication initiation may be designed to be decoupled from the cell state. Such a scheme may simplify the task of engineering global regulation mechanisms such as the one responsible for rRNA regulation in different growth conditions or growth phases.

## Assumptions and Further Predictions

The CCGR models rely on many assumptions, the validity of which should be questioned. One possibility is that the coarse-graining has discarded "hidden variables". Such variables may include, for example, the structure of the nucleoid and transcription factors associated with it (which can affect global transcription [97]), or the osmotic response of the cell [82]. In addition, strong genetic perturbations may lead to ribosome instability [98] and possibly induce a stress response with global effects. Other concerns may be possible additional factors regulating rRNA synthesis alluded to earlier, the validity of the assumptions regarding the function of the bulk protein and the existence of limiting resources even in a rich environment. In a resource limited environment for example, state variables related to the energy metabolism of the cell would probably come into play. Although, regarding limitation of resources, as was pointed out in the Introduction, it has been demonstrated experimentally that the concentration of NTP is constant or changes by only a small amount when altering the rRNA operon copy number [20,64], and ppGpp is also constant in these strains [20,54]. The latter observation suggests that the cell is not limited, for example, by the availability of amino acid, charged tRNAs or carbon [69,89] (see also [6]). Another concern may be that some portion of the inactive RNAp, which was assumed to be inaccessible because of pausing, is actually nonspecifically bound to DNA [28] and might serve as an additional reservoir of RNAp for transcription initiation. With all these difficulties in mind, the advantage of the CGGR modeling approach is that it offers an initial conceptual framework for thinking about *E. coli* while making quantitative predictions. Such tests can be useful in identifying factors that have been left out in this round of coarse-graining and can be subsequently added. Examples of quantitative predictions include: (i) non-constancy of the macromolecular volume fraction in genetically perturbed cells (Figure 3, insert) (ii) state variables and their relations, e.g. the cost criterion (Figure S5) (iii) decay of binding affinities at high volume fractions (Figure 3 and Figure S8; raising the more general question of the nature of crowding effects on equilibrium constants) (iv) increase in bulk mRNA half-life with rRNA operon copy number. Yet another test to this model may be to increase rRNA gene dosage beyond the WT gene dosage, where the differences between the CGGR models is much more pronounced [28] (Figure 2B). Although the focus here was on altering the rRNA operon copy number, other genetic perturbations can be considered, like adding non-native proteins that only serve as a load on the cell. In such a case, *in vivo* diffusion times are expected to be increased due to increased crowding. Green fluorescent protein (GFP) diffusion coefficient did in fact appear to decrease in *E. coli* cells overexpressing GFP, however GFP dimerization may have contributed to this effect, as noted by Elowitz et al. [99]. Finally, the proposed model may suggest testable predictions for the effect of genetic noise on protein expression and growth rate.

## Supporting Information

**Text S1**  Complete Supporting Information

**Table S1**  Genetic parameters for *E. coli* growing at 1 and 2.5 doub/h, 37°C.

**Table S2**  Cell state and additional parameters for various growth conditions.

**Table S3**  Reconstruction of the WT cell state for 1 and 2.5 doub/h, 37°C.

**Table S4**  Transcription related parameters for 2 doub/h, 37°C.

**Table S5**  Genetic parameters for *E. coli* growing at 2 doub/h, 37°C.

**Table S6**  Lineage of the *rrn* inactivation strains.

**Table S7**  Genetic parameters for the *rrn* inactivation strains at 2 doub/h, 37°C.

**Figure S1**  Mean square errors with respect to the Squires data. (A) Unconstrained CGGR MSE. Square root of the mean square error (MSE) as a function of $c_{ribo}$ in estimation of the growth rate and the rRNA to total protein ratio measured by Asai et al. [19]. This graph was computed as follows: for a given $n_0$, optimal $L_{m,bulk}$ and $c_{ribo}$ that minimize the square error between an estimated WT cell state and the observed WT cell state were obtained (see S1.1.1 in Text S1). Next, for those optimal $L_{m,bulk}$ and $c_{ribo}$ values, the growth rate curve and the rRNA/total protein curve were calculated for the various *rrn* inactivation strains (c.f. S1.3 in Text S1) and the MSEs were calculated between these two curves and the data points, yielding two errors for a given $n_0$ (or equivalently $c_{ribo}$). Next, $n_0$ is increased and the process is repeated. The minimum MSE for the rRNA to total protein ratio (which displayed more sensitivity to $c_{ribo}$ than the growth rate) was obtained for $c_{ribo} = 37.6$ ($n_0 = 2.8 \cdot 10^6$ molec/WT cell). Circles mark the cost for which $\Phi$ would be fixed in an unconstrained CGGR model (i.e. when $c_i = -v_i/v_{bulk}$, which is equivalent to the constrained CGGR model with $h=0$). (B) Constrained CGGR MSE. Square root of the MSE in estimation of the growth rate and the rRNA to total protein ratio as a function of $M_{bulk}$ and the Hill coefficient $h$, for a model where $\Phi$ is assumed to be fixed, and $c_p = c_p^{max} / \left[ 1 + (M_{bulk}/n_{bulk})^h \right]$. This graph was computed as follows: for a given $c_p^{max}$ and $h$, optimal $L_{m,bulk}$ and $M_{bulk}$ that minimize the square error between the estimated WT cell state and the observed WT cell state were obtained. Note that this square error included the error between the estimated WT $c_p$ and the observed WT value of $c_p$ at 2 doub/h (20 aa/sec). The error in prediction of the WT cell state was on the order of a few percent (data not shown). Next, for those optimal $L_{m,bulk}$ and $M_{bulk}$ values, the growth rate curve and the rRNA/total protein curve were calculated for the various *rrn* inactivation strains and the MSE was calculated between these curves and the data points. Next, $c_p^{max}$ is increased and the process repeated. The minimum Hill coefficient to yield a solution that did not diverge in growth rate for high *rrn* copy numbers was $h=2$ (see e.g. Figure S2 for fit with $h=1$). For $h=2$, $M_{bulk}$ was chosen to minimize the growth rate error yielding: $M_{bulk} = 7.4 \cdot 10^6$ molec/WT cell ($c_p^{max} = 73$ aa/sec). Solutions that minimized the rRNA/total protein MSE (corresponding to the minimum possible value for $c_p^{max}$, i.e. $\cong 21$ aa/sec) diverged in growth rate for copy numbers greater than 7 (see Figure S3). In addition, the MSE did not improve for higher Hill coefficients, as shown. Note that the minimization procedure in (A) and (B) are equivalent if we map $M_{bulk} \leftrightarrow c_{ribo}$, $c_p^{max} \leftrightarrow n_0$. (C) Simplified 3-state model MSE. Square root of the MSE in estimation of the growth rate and the rRNA to total protein ratio as a function of $c_{ribo}$ for the simplified model. Stars indicate minima. Circles indicate the same as in (A). The minima almost coincide and were obtained for $c_{ribo} \simeq 38.2 \pm 2.8$. In both (A) and (B), as in Figure 2 to Figure 4, $U_{bulk}^{max}$ was set to 80 ini/min and the rRNA chain elongation rate, $c_{rrn}$, was assumed to be constant.

**Figure S2**  Fit for the constrained CGGR model with Hill coefficient $h=1$. Comparison of the constrained CGGR model with Hill coefficient $h=1$ to (A) growth rate measurements and (B) rRNA to total protein ratio measurements of Asai et al [19]. $M_{bulk}$ was chosen such that the product of growth rate error and rRNA to total protein error was minimal, yielding $M_{bulk} = 5.7 \cdot 10^6$ molec/WT cell ($c_p^{max} = 45$ aa/sec). For MSE see Figure S1B. Note that for $h=1$, growth rate diverges with copy number. rRNA chain elongation rate, $c_{rrn}$, was assumed to be constant in this simulation.

**Figure S3**  Fit for the constrained CGGR model with higher Hill coefficients. Comparison of the constrained CGGR model with Hill coefficients of 2, 4, 6, 8, and 10 to (A) growth rate measurements and (B) rRNA to total protein ratio measurements of Asai et al [19]. We show the $h=2$ case for both $c_p^{max} = 73$ aa/sec ($M_{bulk} = 7.4 \cdot 10^6$ molec/WT cell; as in Figure 2) and $c_p^{max} = 21$ aa/sec. For all other cases, $c_p^{max}$ was set to 21 aa/sec and corresponds to the minimum possible value for $M_{bulk}$, a value that according to Figure S1B minimizes the MSE for the rRNA to total protein ratio. This figure demonstrates that all solutions with $c_p^{max} = 21$ aa/sec diverge in growth rate for *rrn* copy numbers greater that 7. Higher Hill coefficients ($>10$) appear to be numerically unstable or insolvable for high copy numbers. Legend to both figures is given in (A).

**Figure S4**  Free RNAp and free ribosomes with respect to corresponding binding affinities for various crowding scenarios. (A) Model prediction for $n_{RNAp,free}$, $n_{RNAp,free}/K_{m,i}$ and $\Phi$ for the transition state limited and no crowding scenarios as a function of the *rrn* operon copy number. In the no crowding scenario the plots for $n_{RNAp,free}$ and $n_{RNAp,free}/K_{m,i}$ coincide. (B) Same as (A) but for the diffusion limited scenario. (C) Model prediction for $n_{ribo,free}$ and $n_{ribo,free}/L_{m,i}$ for the transition state limited and no crowding scenarios as a function of the *rrn* operon copy number. (D) Same as (C) but for the diffusion limited scenario. All curves are normalized to WT values at copy number 7. Note that in the diffusion limited scenario, when rRNA operons are inactivated, free RNAp concentration actually decreases. The reasons for this are that first, although the rRNA operons are inactivated, they continue to be partly transcribed (c.f. S1.3 in Text S1). Second, as rRNA operons are inactivated, growth rate is reduced (Figure 2A), which tends to slightly increase gene concentrations via Eq. 3 (c.f. Figure S7B). Finally, there is the contribution of increased transcription initiation. When rRNA operons are increased beyond seven copies per chromosome, free RNAp concentration increases mainly because transcription initiation is reduced due to diminished

binding affinities. See main text and S1.6 in Text S1 for further explanations.

**Figure S5** Predictions for bulk protein and ribosome concentrations as a function of the *rrn* operon copy number. (A) Total concentration of ribosomes (ribosomes per unit volume) in the constrained and unconstrained CGGR models as a function of the *rrn* operon copy number. (B) Concentration of bulk protein (proteins per unit volume) in the constrained and unconstrained CGGR models as a function of the *rrn* operon copy number. Solid lines are for fixed *rrn* chain elongation rate, $c_{rrn}$ = const, and dashed lines are for $c_{rrn} \neq$ const, as described in the main text. All curves are normalized to WT cell state values (at copy number = 7).

**Figure S6** Breakdown of the ribosome synthesis equation to components for the diffusion limited scenario. (A) Variables in units of concentration. $d_{rrn}$ - *rrn* gene concentration (total rRNA operon copy number per unit volume); $V_{rrn}$ - *rrn* initiation rate per operon (init/min/operon); $n_{ribo}$ - ribosome concentration (ribosomes per unit volume), and μ - growth rate. These parameters are tied by Eq. 2iii: $\alpha = d_{rrn} \cdot V_{rrn}/n_{ribo}$. (B) Variables in units of molec/cell. $D_{rrn}$ - *rrn* gene dosage (total rRNA operon copy number per cell); $N_{ribo}$ - number of ribosomes per cell. These parameters are tied by Eq. 2iii: $\alpha = D_{rrn} \cdot V_{rrn}/N_{ribo}$. This simulation is for the diffusion limited scenario assuming that the rRNA chain elongation rate, $c_{rrn}$, is variable, as described in the main text. All curves are normalized to WT cell state values (at copy number = 7).

**Figure S7** Gene dosage and gene concentration as a function of growth rate. (A) Gene dosage and (B) gene concentration for the *rrn* gene class and bulk gene class. C and D periods were interpolated based on data from table 2 of [17] as a second order polynomial in $\mu^{-1}$. For this simulation we assumed that 66 evenly distributed bulk genes are expressed (c.f. map locations in Table S1). The initiation volume, $V_{ini}$, was assumed to be fixed [41,43,100]. See also main text and S2.2 in Text S1 for further explanations.

**Figure S8** Dependence of binding affinities on the volume fraction Φ for the various crowding scenarios. (A) Normalized inverse equilibrium constants, $K_m^{-1}$ and $L_m^{-1}$ (in units of 1/M), for the RNAp holoenzyme (radius 5.57 nm) and the 30S ribosome subunit (radius 6.92 nm), respectively, in the transition state limited model. The water molecule radius was taken to be 0.138 nm [101] and the radius of the background crowding agent was taken to be 3.06 nm [46]. (B) Normalized $K_m^{-1}$ and $L_m^{-1}$ for the diffusion limited model (curves overlap). All curves were normalized to values at the WT volume fraction of Φ = 0.34. See S2.4 in Text S1 for more details.

## Acknowledgments

We thank Uri Alon, Rob Phillips, Carlos Bustamante, Shalev Itzkovitz, Irene A. Chen, Michael L. Shuler, Evgeni V. Nikolaev, Gideon Schreiber, Maarten H. de Smit, Moselio Schaechter, Hans Bremer, Tomer Kalisky, Alon Zaslaver, Erez Dekel, David Wu, Elisha Moses, Roy Bar-Ziv, and Ofer Vitells for their insightful comments. We are especially grateful for the assistance of Catherine L. Squires and Selwyn Quan.